# ROUTING ALGORITHM FOR SOFTWARE DEFINED NETWORK BASED ON BOX-COVERING ALGORITHM

Dana Turlykozhayeva
*Department of Solid State Physics and Nonlinear Physics*
*Al-Farabi Kazakh National University*
Almaty, Kazakhstan
turlykozhayeva.dana@kaznu.kz

Sayat Akhtanov*
*Department of Solid State Physics and Nonlinear Physics*
*Al-Farabi Kazakh National University*
Almaty, Kazakhstan
ahtanov.saiyat@kaznu.kz

Nurzhan Ussipov
*Department of Solid State Physics and Nonlinear Physics*
*Al-Farabi Kazakh National University*
Almaty, Kazakhstan
ussipov.nurzhan@kaznu.kz

Almat Akhmetali
*Department of Solid State Physics and Nonlinear Physics*
*Al-Farabi Kazakh National University*
Almaty, Kazakhstan
akhmetali_almat@live.kaznu.kz

Aslan Bolysbay
*Department of Solid State Physics and Nonlinear Physics*
*Al-Farabi Kazakh National University*
Almaty, Kazakhstan
bolysbay_aslan@live.kaznu.kz

Yerkin Shabdan*
*Department of Intelligent Systems and Cybersecurity*
*Astana IT University*
Astana, Kazakhstan
y.shabdan@astanait.edu.kz

*Abstract* — A routing algorithm is the most fundamental problem in complex network communication. In complex networks, the amount of computation increases as the number of nodes increases which reduces routing performance. In this paper, we propose a routing algorithm for software-defined networking (SDN) based on a box-covering (BC) algorithm. It is known that using the BC algorithm it is possible to increase performance in complex SDN. We partition the entire SDN network into subnets using three existing box-covering methods such as MEMB, GC and CIEA, then we use Dijkstra's algorithm to find the shortest path between subnets and within each subnet. We compared all box-covering algorithms and found that the GC algorithm has the highest performance for SDN routing.

## I. Introduction

The infrastructure of the traditional network, which is no longer able to keep up with the rapid development of the Internet, increases the operational complexity and cost of the network. An emerging networking technology called Software-Defined Networks (SDN) aims to virtualize the server and storage infrastructures of modern data centers while also making networks flexible and agile. Also, SDN can control, initialize, modify, and manage the network behavior programmatically. To communicate with applications and networking devices, the SDN Controller uses conventional API (Application Programming Interface) in both the northbound and southbound directions. Based on a global network view, the SDN controller calculates routing tables between nodes. [1-4].

The routing mechanism is a crucial component of network resource optimization technology, but due to inherent limitations in traditional networks and the uncontrollability of distributed routing algorithms, it is very challenging to achieve optimal performance. The properties of SDNs listed above offer a new perspective on network routing. SDN's controller can efficiently manage network resources and enable network traffic control.

Additionally, the Dijkstra algorithm is one of the well-known routing algorithms for finding the shortest path, and it is not suitable for determining the shortest past in huge complex networks, like data centers, which can contain millions of nodes. The box-covering (BC) algorithm is used to analyze the fractal dimension of complex networks by covering the network with the minimum possible number of boxes. Moreover, the use of this approach in routing provides renormalization in complex networks and helps to improve its performance [5-11].

In this article, we use box-covering algorithms to improve the performance of a routing protocol for SDN.

Firstly, we divide the SDN network into many subnets using three box-covering methods, namely maximum excluded mass burning (MEMB) [6], greedy coloring (GC) [10], and center-including eccentricity algorithm (CIEA) [11]. Then Dijkstra's algorithm is applied to determine the shortest path between subnets and within each subnet.

Secondly, we calculated and compared the performance of the proposed box-covering based routing algorithm for finding the shortest path. Through this estimation, we revealed with which of the above-mentioned box-covering algorithms the routing of SDNs can be optimized.

The article is organized as follows. In section 1, we discuss related work problems. In section 2, we present details of our proposed algorithm, including the box-covering algorithms used in our study. In section 3, we present and discuss our results. Finally, in section 4 we summarize our results, followed by some suggestions for future work.

## II. Related work

In complex networks, the transportation of data packets is executed through multi-hop routing algorithms. The effectiveness of these routing algorithms largely depends on how they choose the next node to send the packet.

Numerous algorithms can be applied for determining the shortest path in network routing. However, these algorithms also have certain limitations. One such limitation is the need for real-time computing to determine the most efficient routes with minimal resource consumption. This process can be dynamic and non-linear, adding complexity to the calculations. As the number of nodes in a network increases, the computational cost of running routing algorithms also increases, which can result in the production of unnecessary or redundant information. A second major challenge with traditional networks is the difficulty in obtaining global information. For example, algorithms such as Routing Information Protocol (RIP) and Open Shortest Path First (OSPF) come with several problems. Firstly, the performance required to build paths may be too long [12-13]. Secondly, the instability of the prefix may also lead to route map problems [14]. Conflicting routing among autonomous systems can also destabilize the entire network [15]. Finally, the deployment of routing in a large network is a complex process [16].

The development of SDNs is supported by three crucial technologies: programmable network functions within the network, separation of the data forwarding plane and control plane, and network virtualization [17-21]. In the early stages of SDN protocols, the focus was on control plane programmability. However, recent SDN protocols support a broader range of data plane functions. The functionality of network devices in an SDN is managed by software, allowing a network engineer to reprogram network infrastructure as required.

The control plane of SDNs can be physically separated from the data forwarding plane. An SDN network that is disconnected from the main physical equipment greatly simplifies the managing and operating of complex SDNs [22-23].

With the availability of SDN technology, testing and implementing new routing algorithms have become easier. The programmability of SDN has allowed the development of new routing algorithms. Some studies have extended Dijkstra's shortest path algorithm to consider both edge weights and node weights for a graph derived from the SDN topology. Additionally, there have been analyses of the suitability of different routing algorithms for setting up performance-guaranteed traffic tunnels in large-scale SDN networks.

In SDN environments, there are routing algorithms that utilize multipaths to decrease the routing load. Additionally, some routing algorithms dynamically adapt their parameters based on real-time link information. However, these algorithms do not consider the network topology information [24-25].

### III. PROPOSED ROUTING ALGORITHM AND USED BOX-COVERING ALGORITHMS

The existing routing algorithms for SDNs primarily rely on multipath routing and dynamic link calculations. In this paper, we can optimize the routing strategy by considering the network topology information and by reducing the number of nodes and edges in the network, resulting in a lower complexity of the Dijkstra algorithm. The proposed box-covering based routing algorithm is considered even more optimal and simpler than the Dijkstra algorithm. Here we present a step-by-step description of our method:

Firstly, the BC algorithm is used to divide the entire network into multiple subnets, with each subnet being covered by a box of size $l=(2 r_b+1)$, where $r_b$ is the radius of a box.

Secondly, each subnet is treated as a new node, and the Dijkstra algorithm is applied to determine the shortest path between these new nodes.

Thirdly, we find the most efficient route between nodes within each subnet by the Dijkstra algorithm.

In the final step, we connect the shortest path between subnets and the shortest path within subnets and create an overall efficient routing.

Box-covering algorithms were originally developed to calculate the fractal dimension of complex networks. One of the most appealing aspects of complex networks is that they analyze problems based on their structure, which is particularly useful for understanding large-scale networks. The BC algorithm for its implementation requires two parameters as input: a network graph G and a specific value for the box size, denoted as l. Varying the box size l can result in different outputs for the shortest path. In the article, we used 2 common box-covering algorithms in SDN routing such as GC, MEMB, and 1 algorithm developed recently, namely the CIEA. Next, we are going to briefly describe them.

The GC algorithm is widely used in network analysis and has proven to be effective in identifying community structures in a variety of networks, including social networks, biological networks, and transportation networks. The GC algorithm can be broken down into the following steps. First, the algorithm creates a dual network G' to approximate the optimal solution for any given value of $r_b$. This is achieved by connecting two nodes in G' if their chemical distance in the original network G is greater than or equal to $r_b$. After setting up the initial steps, the aim is to assign colors to the nodes of the dual graph in such a way that neighboring nodes cannot have the same color and use as few colors as possible in the coloring process. In figure 1 we present the implementation of the GC algorithm using a model network G [10].

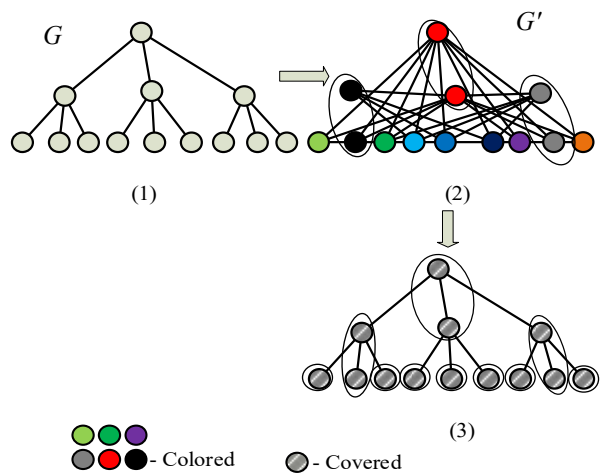

Fig. 1. The implementation of the GC algorithm ($r_b$ = 1) using a model network G.

The MEMB method guarantees the connectivity of boxes by choosing a node that has the maximal excluded mass as the center. The MEMB algorithm consists of the following steps. First, the algorithm begins by selecting a center node based on the maximal excluded mass, which is the number of uncovered nodes within a radius $r_b$ from the center. In the second step, the next central node is selected and all uncovered nodes within $r_b$ from the central node are covered. Finally, the algorithm proceeds to select the next center node by repeating the process until all nodes are covered. In figure 2 we present the implementation of the MEMB algorithm, using a model network G [6].

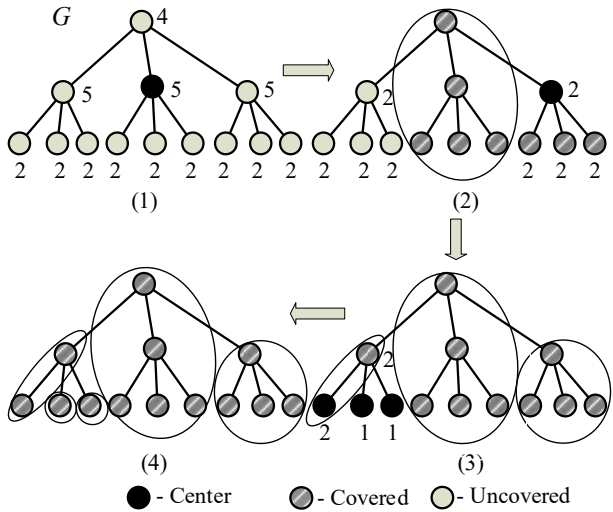

Fig. 2. The implementation of the MEMB algorithm ($r_b = 1$) using a model network G.

The CIEA algorithm has a similar structure to the MEMB method, but it is capable of solving the cases with single-node boxes at the edges of the network. The CIEA does not require complex calculations and is relatively simple to implement. The CIEA consists of several steps. First, the eccentricity of the nodes and the central node is calculated. The central node is chosen to be the node that is situated at a distance $r_b$ from the node with the maximum eccentricity. In the second step, nodes located at a distance $r_b$ from the central node are grouped into one box, which we call in this article a subnet. The process is repeated until the entire network is covered. In figure 3 we demonstrate the implementation of the CIEA using a model network G [11].

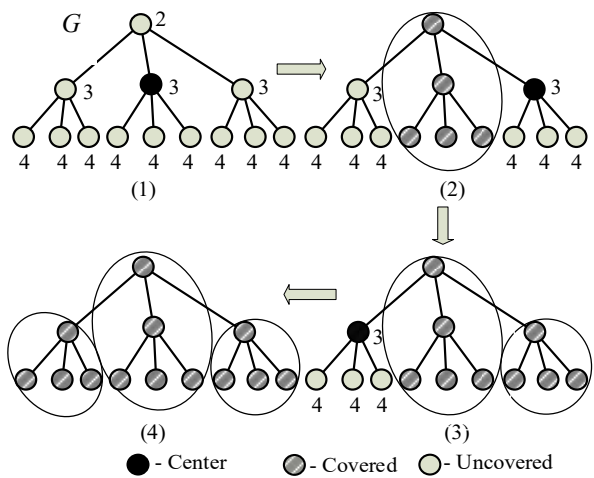

Fig. 3. The implementation of the CIEA ($r_b = 1$) using a model network G.

The BC algorithm is a technique used to restructure large-scale networks in such a way that it reduces the number of nodes and edges in the network, while still preserving the network's essential structural characteristics and functions. The routing process is linked to the network layer, which implies that the routing algorithm used for SDNs of a significant scale is determined based on the topology structure [10-11],[26].

IV. RESULTS AND DISCUSSIONS

To evaluate the performance of our routing algorithm for different BC (GC, MEMB, CIEA), we performed and compared several applications, by running the proposed algorithm on the SDN model of different sizes.

Figure 4 below shows the dependence of the number of boxes on the network size for different BCs.

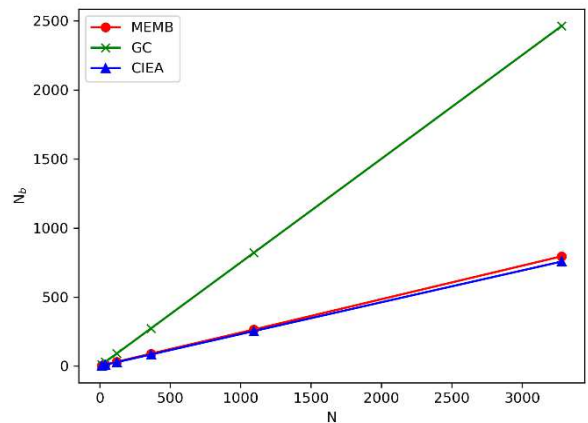

Fig. 4. Dependence of the number of boxes on the size of the network ($r_b = 1$)

According to figure 4, GC-based routing demonstrates a larger number of boxes than the MEMB and CIEA-based routing algorithms. The performance of our box-covering based routing algorithm improves as more boxes are created as a result of division using the BC algorithm.

After that, we compared the BCR algorithms among our algorithm and with the Dijkstra algorithm to determine the best algorithm for dividing into boxes. The results of the comparison by the running time using 3 different BC algorithms and the Dijkstra algorithm are shown in table 1 and in Figure 5.

The results of the running time (in microseconds).

| Number of Nodes | Dijkstra algorithm | BCR (MEMB) | BCR (Greedy Coloring) | BCR (CIEA) |
|---|---|---|---|---|
| 13 | 4.68 | 14.6 | 9.67 | 15.5 |
| 39 | 7.89 | 16.8 | 10.2 | 16.2 |
| 121 | 13.1 | 22 | 10.5 | 25.3 |
| 364 | 22.1 | 25.7 | 14.3 | 33.2 |
| 1093 | 37.3 | 28.5 | 18.0 | 33.5 |
| 3280 | 61.5 | 36.8 | 25.2 | 42.2 |

Table 1 presents the node number in the first column, while the second and subsequent columns indicate the time taken by Dijkstra's algorithm and other routing algorithms based on various BC algorithms to find the shortest path. To facilitate comparison, we have assumed that the size l of each box is 1.

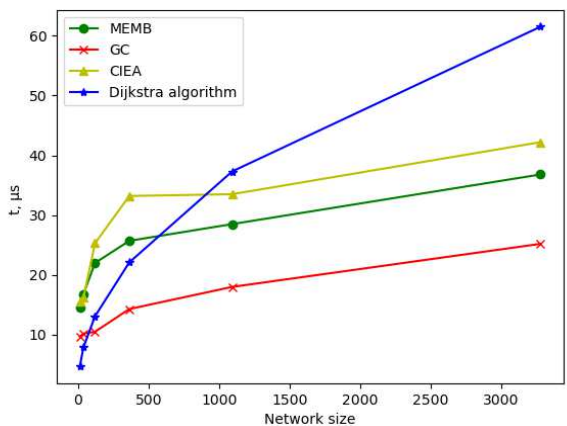

Fig. 5. The implementation of the CIEA ($n_b = 1$) using a model network $G$

As shown in figure 5, the advantages of the BCR algorithm become more obvious as the number of nodes increases rapidly. The Dijkstra algorithm, while being a reliable algorithm, becomes significantly slower as the number of nodes increases, making it less practical for larger SDN networks. When the node number is 3280, the running time of GC based routing algorithm is significantly less compared to when we used MEMB and CIEA algorithms in SDN routing.

## V. CONCLUSION

Traditional network routing algorithms are often complex and difficult to control, resulting in low utilization of network resources. However, the emergence of SDN technology provides a promising solution for improving the usage rate of network resources. SDN separates control from the data forwarding plane and allows you to implement many routing algorithms that may be difficult to implement in traditional networks. This paper proposes a BCR algorithm through the modeling of the SDN in a Python environment. There are two approaches to current SDN routing algorithms, multi-path routing and the other involves calculating the route according to the real-time renewed information. Traditional routers can only acquire local topology information to route traffic, but SDN's central controller can acquire global topology information. Our proposed algorithm is based on topology information, and it efficiently divides the whole network into many logical subnets using box-covering methods. The validity of the BCR algorithm has been proven by several examples, and as a result, the proposed algorithm performs better in large-scale networks than in others. Also, we revealed that using the GC box-covering algorithm in SDN routing can significantly enhance the network performance in terms of finding the shortest path in the shortest time compared to when we used MEMB and CIEA in SDN routing.

ACKNOWLEDGMENT

We would like to express our sincerest gratitude to the Al-Farabi Kazakh National University for supporting this work by providing computing resources (Department of Physics and Technology). This research was funded by the Committee of the Ministry of Science and Higher Education of the Republic of Kazakhstan, grant AP19674715.

*Equal Correspondence*